\documentclass[letterpaper, 10 pt, conference]{ieeeconf}  

\IEEEoverridecommandlockouts                              
\usepackage{graphics} 
\usepackage{graphicx}
\usepackage{physics}
\usepackage{caption}
\usepackage{subcaption}
\usepackage{enumerate}
\usepackage{float}
\usepackage{mathtools}
\usepackage{multirow}
\usepackage{amssymb}
\usepackage{amsmath}
\usepackage{epstopdf}
\usepackage[space]{cite}
\newtheorem{definition}{Definition}[section]
\title{\LARGE \bf Reinforcement Learning for Test Case Prioritization}

\author{Lousada, João \\
Instituto Superior Técnico, Universidade de Lisboa \\
joao.b.lousada@tecnico.ulisboa.pt
\and
Ribeiro, Miguel \\
BNP Paribas  \\
miguel.a.ribeiro@tecnico.ulisboa.pt
}

\begin{document}

\maketitle
\thispagestyle{plain}
\pagestyle{plain}
\vspace{-1cm}

\begin{abstract}

In modern software engineering, Continuous Integration (CI) has become an indispensable step towards systematically managing the life cycles of software development. Large companies struggle with keeping the pipeline updated and operational, in useful time, due to the large amount of changes and addition of features, that build on top of each other and have several developers, working on different platforms. Associated with such software changes, there is always a strong component of Testing. As teams and projects grow, exhaustive testing quickly becomes inhibitive, becoming adamant to select the most relevant test cases earlier, without compromising software quality.
This paper extends the studies conducted by Spieker \textit{et al.} \cite{Spieker} on applying Reinforcement Learning to optimize testing strategies. We test its ability to adapt to new environments, by testing it on novel data extracted from a financial institution, yielding a Normalized percentage of Fault Detection (NAPFD) of over $0.6$ using the Network Approximator and Test Case Failure Reward.
Additionally, we studied the impact of using Decision Tree (DT) Approximator as a model for memory representation, which failed to produce significant improvements relative to Artificial Neural Networks.

\vspace{0.25cm}

\end{abstract}

\vspace{0.25cm}

\section{Introduction}

\textbf{Context} Given the complexity of modern software systems, it is increasingly crucial to maintain quality and reliability, in a time-saving and cost-effective manner, especially in large and fast-paced companies. This is why many industries adopt a \textit{Continuous Integration} (CI) strategy, a popular software development technique in which engineers frequently merge their latest code changes, through a \textit{commit}, into the mainline codebase, allowing them to easily and cost-effectively check that their code can successfully pass tests across various system environments \cite{santolucito2018statically}.

\textbf{Regression Testing} One of the tools used to manage software change is called regression testing. It is critical to ensure that the introduction of new features, or the fixing of known issues, is not only correct, but also does not obstruct existing functionalities. Regressions occur when a software bug causes an existing feature to stop functioning as expected after a given change and can have many origins (e.g. code not compiling, performance dropping, etc.), and, as more changes occur, the probability that one of them introduces a fault increases \cite{ShinThesis}.  

As software development teams grow, identifying and fixing regressions quickly becomes one of the most challenging, costly and time-consuming tasks in the software development life-cycle, rapidly inhibiting its adoption. Such teams often resort to modern large-scale test infrastructures, like core-grids or online servers \cite{Ziftci}. Consequently, in the last decades, there has been intensive research into solutions that optimize Regression Testing, accelerating fault detection rates either by alleviating the amount of computer resources needed or by reducing feedback time, i.e. the time delay between a software change and the information regarding whether it impacts the system's stability \cite{palma, Uber, yangJIT, learningfortcp, liang}.

The most prominent techniques for Regression Testing optimization are Test Case Minimization, where the number of tests is trimmed to avoid redundancies, Test Case Selection, where only a subset of all the tests is chosen, and Test Case Prioritization (TCP), where more relevant test cases are run first. This has become one thriving field, proven to have achieved meaningful results, with increasing research attention \cite{durelli, litrevtcp}.

More specifically, TCP aims to find the optimal permutation of test cases that matches a certain target, e.g. the ability to reveal faults as soon as possible, which is useful when there's a time budget or computer resources are limited \cite{ShinThesis}. The key goal of this study is to find out, \textit{a priori}, which test cases to prioritize, i.e. predicting which test cases will fail given a set of changes in the codebase. One possible solution would be to have a professional test engineer cherry-pick the most promising test cases. Unfortunately manual test case selection is time-consuming, counter-productive and error-prone, and is not scalable \cite{durelli}. Therefore, there has been a high demand for techniques that can automatically select test cases, minimizing human intervention \cite{Ziftci}. 

\textbf{Reinforcement Learning}
Reinforcement Learning (RL) handles problems that involve learning which course of action to take, given a set of possible states and possible actions. Each action taken produces a given reward.  In RL, the goal of an \textit{agent}, i.e. the decision maker, is to interact with the environment and select the actions that maximize the cumulative sum of the reward signal. The agent's ability to design an optimal strategy is strongly dependent on three factors: the way the reward function is defined, which features are fed into the model and its ability to generalize instead of memorize. 
\par When dealing with a large dataset, wherein state space representation complexity grows, it is not feasible to represent the state space in a tabular manner, i.e. store the state space discretely in a table.  Hence, to reduce the memory needed to represent the state space, we use Approximators (described in Section \ref{Approximators} with more detail). These can be  ML algorithms (e.g. Neural Networks, DT's, Nearest Neighbours) and are used as memory representation to accelerate computations \cite{rlintro}.
In the context of TCP, we want the agent to learn how to rank test cases, such that the ones that are more prone to reveal faults have a higher priority than the ones that are not. First, each test case is prioritized individually, so that a test schedule is created, executed and finally evaluated. Traditionally, the only information provided to the algorithm is historical results. In the work done by both Spieker \textit{et al.} \cite{Spieker} and Wu \textit{et al.} \cite{time-window}, the method preferably prioritizes test cases which have been failing recently.

In this paper our contributions are threefold: 

\begin{enumerate}

    \item The exploration of different ML methods to represent state-value functions, such as DT's. 
    
    \item The testing of experimental results against three different industrial data sets, \textit{IOF/ROL}, \textit{PaintControl} and \textit{Finance} (novelty). Thus, we assess whether this method can be generalized, when applied to a different context. 
    
    \item The performance comparison of with traditional prioritization methods.
\end{enumerate}

\textbf{Paper outline} Hereafter, Section \ref{back} will correspond to the Background and Related Work, defining the state of the art. Then, Section \ref{Approximators} presents our approach, explaining how different ML algorithms can represent memory. Section \ref{exp} provides the experimental evaluation of our approach on novel data, in comparison to open-source data sets, as well as threats to validity and future work. Finally, Section \ref{conclusion} summarizes and concludes the paper.

\section{Background and Related Work}\label{back}

\subsection{Software Testing}

In software engineering, version control systems are a means of keeping track of incremental versions of files and documents, allowing the user to arbitrarily explore and recall the past commits that lead to that specific version\cite{santolucito2018statically}. Testing is a verification method used to assess the quality of a given software version. The building block of software testing is the test case, which specifies on which conditions the System Under Test (SUT) must be executed in order to detect a fault, i.e. for a given input, what are the expected outputs \cite{durelli}.

When test cases are applied, the outcome obtained is in the form of PASS/FAIL, with the purpose of verifying functionality or detecting errors. However, testing is very much like sticking pins into a doll - to cover its whole surface a lot of pins are needed, and the larger the doll, the more pins we require. Likewise, the larger and more complex the SUT, the greater the variety of test cases required. Therefore, to ensure that the health of the SUT is maintained throughout time, exhaustive testing is required to cover all possible scenarios \cite{7PrinciplesSoftTest}.

Inevitably, this task becomes impractical or even unfeasible due to the increasing complexity of the SUT, so testers have to find scalable approaches to counteract exhaustive testing, usually resorting to three techniques: Test Case Minimization, Test Case Selection and Test Case Prioritization (TCP), the latter being the target of this work. \\ 

\textbf{Test Case Prioritization}. As mentioned before, TCP rearranges test cases according to a given criteria, such as the probability of revealing failures. It can be formally defined as:

\begin{definition}{\textbf{TCP}}
	Given the set of tests $T$, the set containing the permutations of $T$, $PT$, and a function from $PT$ to real numbers $f : PT \rightarrow \mathbb{R}$, find a subset $T'$ such that
	\begin{equation}
	    [f(T') \ge f(T'')], \hspace{0.5cm} \forall T'' \in PT,
	\end{equation}
	where $f$ is a real function that evaluates the obtained subset in terms of a selected criteria (e.g. code coverage, early fault detection, less resources spent etc.) \cite{ShinThesis}. 
\end{definition}


\subsection{Formalism}

This section provides the necessary formalism for the Test Case Prioritization problem studied throughout this work.
\par The test pool is defined as $T$ and is composed by the set of test cases $\{t_1, t_2, ..., t_N\}$. For each commit $C_i$, a test suite $T_i \subset T$ can be selected and ordered.When there is no time or resource restriction, $T_i$ can be defined to contain the ordered set of the entire test pool, $T_i^{all}$. When such restrictions come into play, a subset of the test pool $T$ is selected for $T_i$. Note that $T$, being the test pool, does not have an ordering, while both $T_i$ and $T_i^{all}$ are meant to be ordered sequences. Therefore, the definition of a ranking function that acts over all test cases should be defined: $ rank : T_i \to N$, where $rank(t)$ is the index of test case $t$ in $T_i$.

Each test case $t$ contains information about its duration, $t.duration_i$, which is known before execution, and its test status, $test.status_i$, only known after execution, which is equal to $1$ if the test has passed, or $0$ if it has failed. In $T_i$, the subset of all failed test cases is denoted $T_i^{fail} = \{ t \in T_i \hspace{0.1cm} s.t. \hspace{0.1cm} t.status_i = 0 \}$ .

\subsection{Machine Learning}

Some problems can not be solved by traditional algorithms, due to limited or incomplete information. In our case, we don't know which tests are more likely to uncover failures. It could be the case, that there is not even a single failure. Furthermore, a test failing previously is not a certain indicator that it will fail again. Hence, with the rise of data availability, there has been a growing interest to investigate solutions that involve learning from data \cite{durelli}.

Benjamin Busjaeger \textit{et al.} \cite{learningfortcp} proposed an optimization mechanism for ranking test cases by their likelihood of revealing failures, in an industrial environment. The ranking model is fed with four features: code coverage, textual similarity, failure history and test age. These features were then used to train a Support Vector Machine model, resulting in a score function that sorts test cases. More recently, Palma \textit{et al.} \cite{palma} trained a Logistic Regression model, fed with similarity based features, such as similar lines of code. Liang \textit{et al.} \cite{liang} proved that prioritization at the commit-level, instead of test-level, would enhance fault detection rates, on fast-paced software development environments.

Another way of achieving effective prioritization is using Semi-Supervised Learning approaches, like clustering algorithms. Shin Yoo \textit{et al.} \cite{Shinyoo} and Chen \textit{et al.} \cite{chen} endorse coverage-based techniques, claiming that making fast pair-wise comparisons between test cases and grouping them in clusters allows for humans to pick, more straight-forwardly, relevant non-redundant test cases, the assumption being that test cases that belong to the same cluster will have similar behaviours i.e. detect the same faults.

%

Recently, Spieker \textit{et al.} \cite{Spieker} were the first to implement a Reinforcement Learning approach to TCP, introducing RETECS. This method prioritizes test cases based on their historical execution results and duration. RETECS has the advantage of adapting to changes in the SUT and new constraints without compromising speed and efficiency. Applied to three different industrial datasets, RETECS challenges other existing methods and has caught attention from other researchers, namely Wu \textit{et al.} \cite{time-window}. This work is an extension of the research conducted by the aforementioned authors.

\subsection{Reward functions}

In a RL problem, the goal consists of collecting numerical rewards that measure the performance of the agent at performing a given task. Hence, properly defining a reward function that reflects these goals will steer the agent's strategy towards optimality. We now use 3 reward functions defined by Spieker \textit{et al.} \cite{Spieker}, namely Failure Count, Test Case Failure and Time-Ranked, defined as


\begin{equation}
    reward_{i}^{fail}(t) = \abs{T_i^{fail}} \hspace{0.5cm} (\forall_t \in T_i),
\end{equation}



\begin{equation}
     reward_{i}^{tcfail}(t) = \begin{cases} 1 - t.status_i, & \mbox{if } t \in T_i \\ 0, & \mbox{otherwise} \end{cases},
\end{equation}



\begin{equation}
     reward_{i}^{time}(t) = \abs{T_i^{fail}} - t.status_i \times \smashoperator{\sum_{\substack{t_k \in T_i^{fail} \wedge \\ rank(t) < rank(t_k)}}} 1,
\end{equation}


where $\abs{T_i^{fail}}$ is the number of failing test cases.




%

\section{Decision Trees Approximators}\label{Approximators}

In the context of TCP, RETECS prioritizes each test case individually and a test schedule is then created, executed and finally evaluated. Each state represents a single test case and it contains information on the test's duration, when it was last executed and the previous execution results. The set of possible actions corresponds to the set of all possible prioritizations a given test case can have in a commit, assigned by the RL agent. After all test cases are prioritized and submitted for execution, their respective rewards are attributed based on the test case status as feedback, assessing performance. From this reward value, the agent can adapt its strategy for future situations: an action yielding positive rewards is reinforced, whereas negative rewards discourage the current behaviour.

When training RL algorithms, in any given state we need to estimate what rewards will be received in the future if each of the available actions is chosen, so we can select the action that maximizes future rewards \cite{rlintro}. These estimates are calculated by defining a \textit{value-function} $v_{\pi}(s)$, where $s$ is the state, with respect to the learned policy $\pi$ and can be learned from experience. 

In tabular representation, there are two options: either the agent finds itself in a new state and registers the obtained value of that state or it finds itself in a state where it had been before and an average of the value is calculated. Multiple encounters of the same state will lead to a more accurate representation of that state's actual value. 
However, with an increasing number of states it may not be feasible to keep track of each state individually. In such a case, $v_{\pi}(s)$ could be a parameterized function, $v_{\pi}(s, \textbf{w})$, where  $\textbf{w} \in \mathbb{R}^n$ is the parameter vector. From now on, we say that $v_{\pi}(s) \approx v_{\pi}(s, \textbf{w})$ for the approximated value of a state $s$ given weight vector $\textbf{w}$ that should be adjusted with experience in order to match observations.

\par These parameterized functions are called \textit{approximators} and, based on the returns of observed states, are able to generalize to unseen ones. This generalization significantly reduces the amount of memory needed to store information about each state, as well as the time required to update it, even though these new values correspond to estimations rather than the actual observed values.
\par The topic of function approximation is an instance of \textit{supervised learning}, because it gathers observed examples and attempts to optimize parameters to construct an approximation of an entire function \cite{rlintro}. A valid example for such a function are Artificial Neural Networks (ANN) where the parameters to be adjusted are the network's weights, which from now on will be referred to as Network Approximator,. The downside of Network is that a more complex configuration is required to achieve higher performance \cite{rlintro}.
\par
\begin{figure}[H]
\centering
\includegraphics[width=0.85\columnwidth]{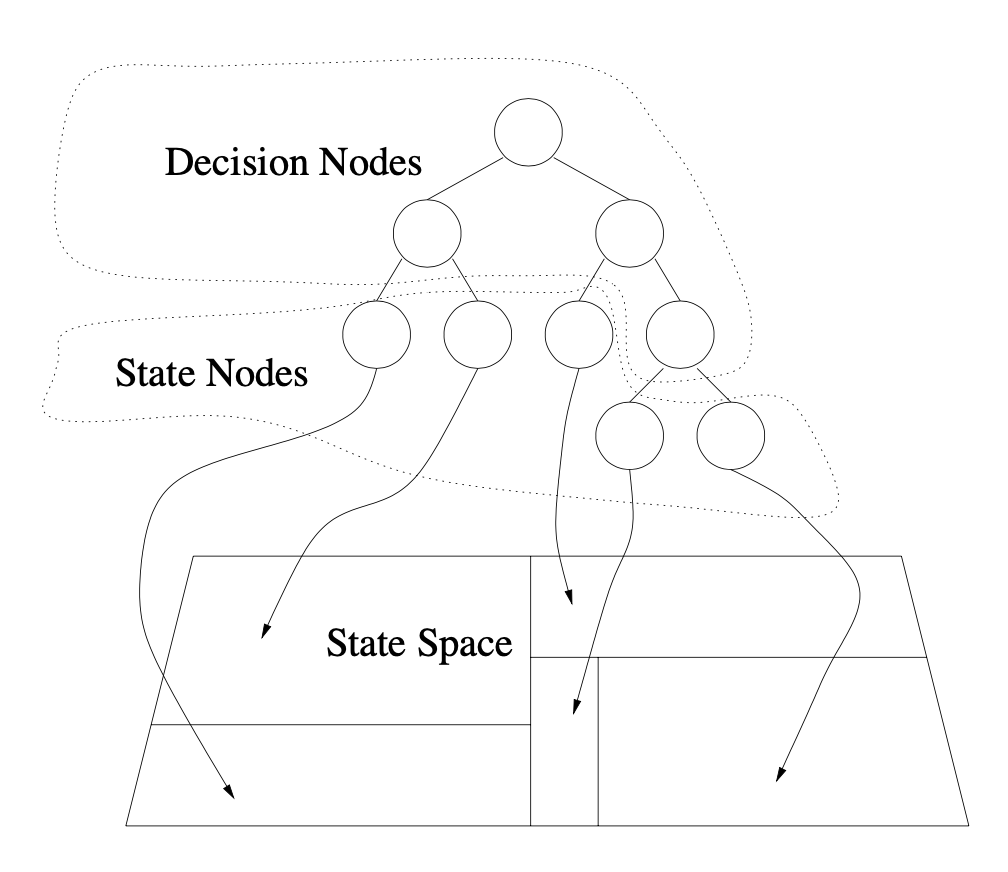}
\caption{Represent state space regions with DT's (Adapted from \cite{dtRL})}
\label{dt}
\end{figure}

 In this work, we implement a DT Agent to approximate the value function. Fig. \ref{dt} shows how a DT can be used to map an input vector, i.e. a state, to one of the leaf nodes that points to a specific region in the state space. The RL agent is then able to learn the values received from taking each path/actions and where they will lead \cite{dtRL}.

By computing the $v_{\pi}(s)$ with a DT, $\textbf{w})$ represents all the parameters defining the splitting points and leaf values of the tree. Commonly, the number of parameters $n$ - length of vector $\textbf{w})$ - is much lower than the number of possible states and tweaking the value of one parameter will affect the value of many states. An action will force the agent to change its state, updating its particular value and consequently, updating $\textbf{w})$, which will generalize from that state to affect the values of many other states. 
\par Finally, the algorithm will generalize better or worse depending on convergence. For example, many supervised learning models' goal is to minimize the root-mean-squared error (RMSE), in this case between $v_{\pi}(s)$ and $v_{\pi}(s, \textbf{w})$ \cite{rlintro}.

\section{Experimental Evaluation}\label{exp}

The next section presents the application and evaluation of our framework, describing, first, the setup procedures as well as a description of the datasets used (section \ref{data}). We then provide an overview of possible evaluation metrics to assess the framework's performance \ref{eval}. To maximize these metrics, fine-tuning is used to find the best combination of parameters, which is shown in section \ref{tune}. Finally, in section \ref{results}, the results obtained are presented and discussed, according to the research questions formulated below. Finally, threats and future work are discussed to close the evaluation process.

\textbf{RQ1}: How will RETECS behave in the presence of a novel dataset with different characteristics? 

\textbf{RQ2}Which function approximator yields better performance? We compare two different models: Artificial Neural Networks and DT's.

\textbf{RQ3}: How is RETECS model performance compared to traditional prioritization techniques, in this new context?

\subsection{Data Description}\label{data}

The data used in this work corresponds to two of the industrial real-world datasets used by Spieker \textit{et al.}  \cite{Spieker} and Wu \textit{et al.}  \cite{time-window}, from ABB Robotics Norway \footnote{Website: http://new.abb.com/products/robotics}, \textit{Paint Control} and \textit{IOF/ROL} which test complex industrial robots. The novel dataset, henceforth referred to as \textit{Finance}, was provided by an investment bank. Each dataset contains historical information of test results, around 300 commits, and has different characteristics. Table \ref{datasets} summarizes the main statistics about the datasets.

\begin{table}[H]
\centering
\resizebox{1.125\columnwidth}{!}{\begin{minipage}{\columnwidth}
\begin{tabular}{lcccc}
\hline
\textbf{Dataset}      & \multicolumn{1}{r}{\textbf{Tests}} & \multicolumn{1}{r}{\textbf{Commits}} & \multicolumn{1}{r}{\textbf{Executions}} & \multicolumn{1}{r}{\textbf{Failed}} \\ \hline
\textit{IOF/ROL}       & 2,086                                   & 320                                  & 30,319                                  & 28.43 $\%$                          \\
\textit{Paint Control} & 114                                     & 312                                  & 25,594                                  & 19.36 $\%$                          \\
\textit{Finance}       & 1,379                                   & 303                                  & 417,837                                 & 63.87 $\%$                          \\ \hline
\end{tabular}
\end{minipage}}
\caption{Dataset Statistics}
\label{datasets}
\end{table}

The datasets are alike in number of commits, but it is clear that the testing strategy is distinct. The \textit{IOF/ROL} dataset contains the least amount of test executions facing the number of test cases it has on the system, meaning that the strategy is much more focused on test case selection. As for the \textit{Finance} dataset, the number of test executions is equal to the number of test cases times the number of commits. Thus, we can conclude that there is no test case selection and every test is applied on each commit and that is why the rate of failed test is higher relative to the other two datasets.

\subsection{Evaluation Metric}\label{eval}

The metric to evaluate the framework's performance is the NAPFD (Normalized Average Percentage of Fault Detection), as defined by Spieker \textit{et al.} \cite{Spieker} and it represents the most common metric to assess the effectiveness of test-case prioritization techniques together with test case selection, that occurs when there is a time limit associated with testing. It is defined as

\begin{gather}
    \text{NAPFD}(T_i) = p - \frac{\sum\limits_{t \in T_i^{fail}} rank(t)}{\abs{T_i^{fail}} \times \abs{T_i}} + \frac{p}{2 \times \abs{T_i}}, \\
    \text{with $p = \frac{\abs{T_i^{fail}}}{\abs{T_i^{total fail}}}$},
\end{gather}
where $\abs{T_i}$ is the number of test cases. Furthermore, the higher its value, the higher the quality of the test schedule will be: if equal to $1$, all relevant test are applied in order, in the beginning, and if it equal to $0$, every relevant test is applied at the end of the schedule. 

In this work, similarly to the original authors, a time limit was imposed, corresponding to $50 \%$ of the total time spent if all test cases were applied. 

It should be also noted that NAPDF is commonly used in its unnormalized form, APFD, where there is no test case selection. In this case, it includes the ratio between found and possible failures within the test subset, such that when $p=1$, all possible faults will be detected and NAPFD reduces to APFD. 

\subsection{Fine-Tuning}\label{tune}

\begin{figure}[b]
\centering
\includegraphics[width=\columnwidth]{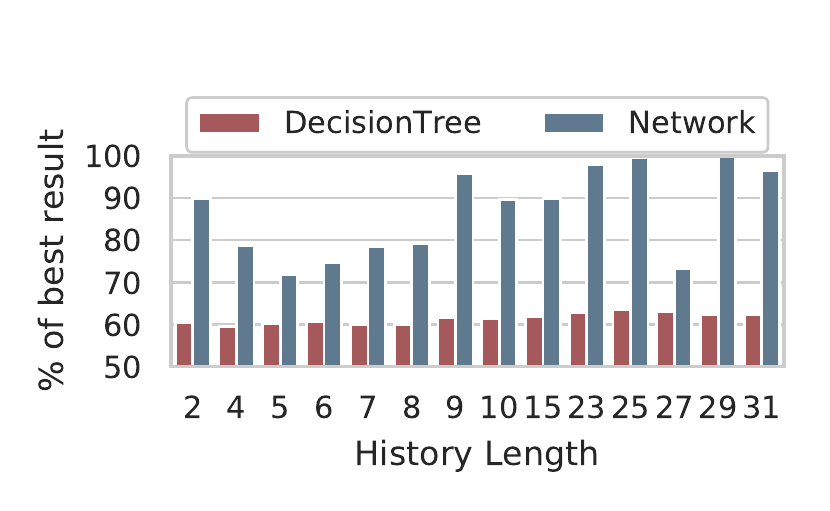}
\caption{History Length Analysis for the \textit{Finance} Dataset for two ML algorithms shown as $\%$ of best APFD. Best Result is 25 executions, corresponding to $100 \%$.}
\label{hlen}
\end{figure}

Parameter tuning allows us to find the best combination of parameters that maximizes the performance of the RL agent, while providing necessary flexibility to adapt to different environments. 
For the \textit{IOF/ROL} and \textit{PaintControl} datasets, the same configuration as Spieker \textit{et al.}  \cite{Spieker} was used, in order to replicate the same results for the Network Agent, which corresponds to using $12$ nodes with one hidden layer. 
\begin{itemize}

    \item \textbf{Network Tune:} For the \textit{Finance} dataset the architecture of the hidden layer for the Network Agent was studied, by calculating the NAPFD with different configurations. A neural networks' architecture may vary in the number of hidden layers, and the number of neurons on each of them.



    \item \textbf{DT Tune}: The parameters to be tuned in DT's are: (1) \textit{criterion}, (2) \textit{maximum depth} and (3) \textit{minimum samples}. (1) measures the quality of a split, where the options are \textit{Gini} for the Gini impurity or \textit{entropy} for the information gain; (2) is the distance between the root node and the leaf node (if depth is infinite the nodes are expanded until all are pure), and (3) is the number of samples needed to split a node in the tree. The variation of these parameters was studied by running a grid search and evaluating the performance for the \textit{Finance} dataset. The results of parameter tuning analysis are summarized in Table \ref{param}. 

    \item \textbf{History-Length} determines how long the execution history should be. A short history-length may not suffice to empower the agent to make meaningful future predictions, although the most recent results are more likely to be more relevant. A larger history-length may encapsulate more failures and provide more fruitful information. However, having a larger history increases the state space and therefore the complexity of the problem, taking longer to converge to an optimal strategy. Moreover, the oldest history record will have the same weight as the most recent, meaning that information about the first test case execution is as relevant as the last execution. Hence, there is no guarantee that a longer execution history will lead to a performance boost. Fig. \ref{hlen} studies how different history length values affect the RL agent, for the \textit{Finance} dataset. 
\end{itemize}

\begin{table}[h]
\centering
\begin{tabular}{clcc}
\hline
\multicolumn{1}{l}{\textbf{RL Agent}}  & \multicolumn{1}{c}{\textbf{Parameter}} & \multicolumn{1}{l}{\textbf{Best Value}} & \textbf{\begin{tabular}[c]{@{}c@{}}Possible \\ Values\end{tabular}} \\ \hline
\multirow{2}{*}{Network} & Nr Layers                              & 1                                       & {[}1,2,3{]}                                                         \\
                              & Nr Neurons                             & 32                                      & {[}12, 32, 64, 100{]}                                               \\ \hline
\multirow{3}{*}{DT}  & Criterion                              & Gini                                    & {[}Gini, Entropy{]}                                                 \\
                              & Min\_Sample\_Split                     & 3                                       & {[}2, 3, 5, 10{]}                                                   \\
                              & Max\_Depth                             & 20                                      & {[}2, 4,8,20{]}                                                     \\ \hline
\end{tabular}
\caption{Parameter Overview}
\label{param}
\end{table}



There does not seem to exist any clear relationship between the two quantities depicted in Fig. \ref{hlen}. The best result obtained corresponds to a history-length of $25$ executions - as well as $29$ but the first one is preferable, so that efficiency is not compromised. This is disparate from the optimal history length obtained by Spieker \textit{et al.} \cite{Spieker} for the other two datasets, which is $4$. This reinforces the fact that \textit{Finance} data has dissimilar characteristics from the other datasets.

\subsection{Results} \label{results}

In our experiments, we trained two RL agents. The first resorts to an Network representation of states, while the second uses a DT. On both cases, the reward function varies between failure count, test-case failure and time ranked. For each test agent, test-cases are scheduled in descending order of priority and until the time limit is reached, if there is one.
Traditional prioritization methods were included as a means of comparison: \textit{Random}, \textit{Sorting} and \textit{Weighting}. 

\begin{itemize}
  \item \textbf{Random} assigns random prioritizations to each test case, and this serves as a baseline method. The other two methods are deterministic. 

  \item \textbf{Sorting} method sorts each test case according to its most recent status, i.e. if a test case failed recently it has higher priority. 

  \item \textbf{Weighting} method is a naive version of RETECS without adaptation, because it considers the same information - duration, last run and execution history - but uses a weighted sum with equal weights. 
\end{itemize}

Due to the fact that RETECS learns incrementally as it is trained, the evaluation metric NAPFD is measured on each commit and due to the exploratory nature of the algorithm, we iterate through the experiments 30 times in order to capture the randomness, averaging out the exploratory nature of the algorithm. Unless stated otherwise, reported results show the mean value of all iterations.
For reproducibility, our contribution to RETECS is implemented in a publicly accessible repository \textit{https://github.com/jlousada315/RETECS}, in Python, using \textit{Scikit-Learn's} toolbox for ANN's and DT's. 

\subsection{RQ1 \& RQ2}

\begin{figure*}[htp]
\centering
\includegraphics[width=\textwidth]{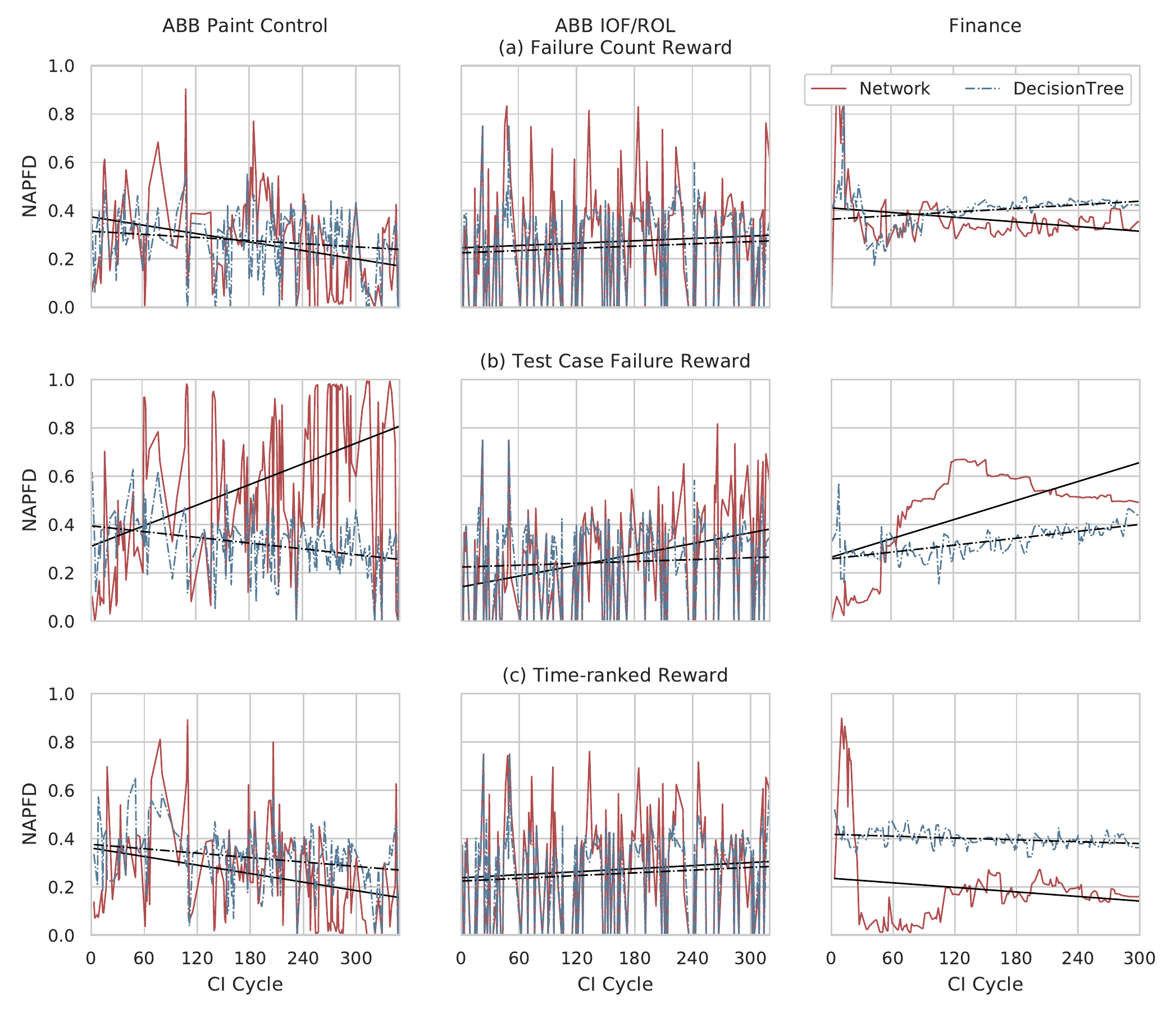}
\caption{NAPFD Comparison with different Reward Functions and memory representations: best combination obtained for Test Case Failure reward and Network Approximator (straight lines indicate trend)}
\label{rq1}
\end{figure*}

Fig. \ref{rq1} shows a comparison of the prioritization performance between the Network Agent and the DT Agent, with regards to different reward functions (rows), applied to three different datasets (columns). The commit identifier is represented in the x-axis and for each one there is a corresponding NAPFD value, ranging from $0$ to $1$. (represented as a line plot in red and blue for the Network and DT agents, respectively). The straight lines show the overall trends of each configuration, which is obtained by fitting a linear function - full line for Network and dot-dashed line for DT Approximator. It is worth noting that for \textit{IOF/ROL} and \textit{Paint Control}, with the Network Approximator, we replicate the results from the original paper \cite{Spieker}, to make sure no errors were introduced while modifying the code.

It is noticeable that both the approximator used for memory representation and the choice of the reward function have a deep impact on the agent's ability to learn better prioritization strategies. For the Failure Count and Time-Ranked reward functions, both approximators go hand in hand and present similar trends, i.e. for a given dataset and reward function, both decrease or increase in approximately the same amount.

However, this behaviour no longer holds when looking at the Test Case Failure reward function. This is the function that evidently produces the best results, in terms of maximizing the slope of the NAPFD trend. When combined with the Network Approximator, this approach proves to be the best configuration overall, for the three datasets, where we see a more significant growth in the trend line, indicating that the algorithm is learning from the data. This implies that attributing specific feedback to all test-cases individually enables the agent to learn how to effectively prioritize them and adapt to heterogeneous environments. 

Another aspect reinforcing the notion of heterogeneity is the differences in the fluctuations of each dataset. For the first two datasets, we can clearly see fluctuations in the results. Spieker \textit{et al.} \cite{Spieker} correlates them with the presence of noise in the original dataset, that may have occurred for numerous reasons and are hard to predict, such as tests that were added manually and produced cascading down failures. Notwithstanding, this behaviour is not observed for the \textit{Finance} dataset, which suggests a much more stable system with less fluctuations, so there is a stronger indication that, with the right set of features, a crystal-clear relation between test cases and their probability of failing can be learned.

In conclusion, the supremacy of the Network Approximator remains valid for the reward function that produces the best results. Yet, in some cases, the DT Approximator was able to surpass its performance by a small amount. If, for example, the \textit{Finance} dataset had more records, it is possible that DT would follow the growing trend and surpass Network by a significant amount. Therefore, the collection of more data is crucial to correctly evaluate the DT Approximator's performance.
\par Choosing the best configuration, test case failure reward and the Network Approximator, when RETECS is applied in an environment completely different from Robotics and with different characteristics, it was able to adapt and learn how to effectively prioritize test cases. This shows that the RETECS domain of validity expands to distinct CI environments, which is particularly useful for companies that increasingly rely on the health and proper functioning of these systems.

\subsection{RQ3}

\begin{figure*}[!t]
\centering
\includegraphics[width=\textwidth]{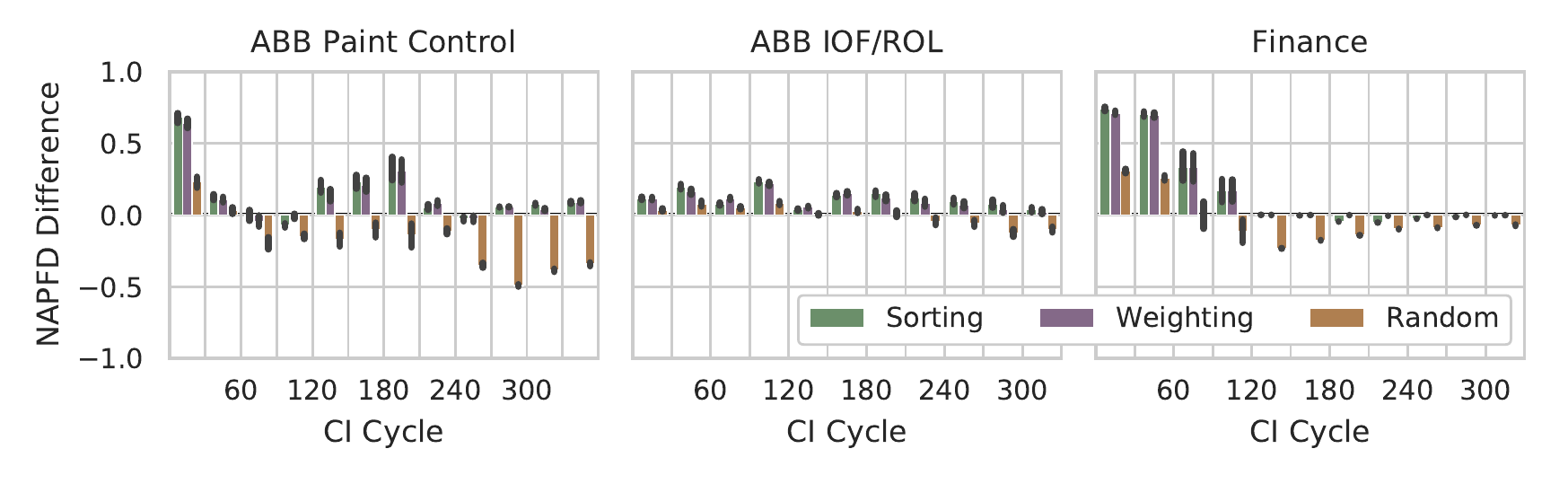}
\caption{NAPFD difference in comparison to traditional methods: Random, Sorting and Weighting. Positive differences indicate better performance from traditional methods and negative differences show better performance for RETECS}
\label{rq2}
\end{figure*}

The focus of RQ1 and RQ2 was to discover which combination of components would maximize performance, which we found to be Test Case Failure Reward and the Artificial Neural Network Approximator. Now, with RQ3, our aim is to compare this approach to traditional test case prioritization methods: \textit{Random}, \textit{Sorting} and \textit{Weighting}. The results are depicted in Fig. \ref{rq2} as the difference between the NAPFD for the traditional methods and RETECS over several CI Cycles. Each bar comprises 30 commits. For positive differences, the traditional methods have better performance, and on the contrary negative differences show the opposite. 
Due to the exploratory character of the algorithm, it is expected that at the beginning, the traditional method will make more meaningful prioritizations and this trend is verified in all datasets, although more evidently in \textit{Paint Control} and \textit{Finance}. 

For the \textit{Paint Control} dataset there are 2 adaptation phases: first, there is a steep convergence in the early commits, with the RETECS method needing only around 60 commits to perform as well as or even better than the traditional methods. Then for the next 90 commits, RETECS performance was progressively worse indicating lack of adaptation and then for the remaining commits, the performances of Sorting and Weighting both match RETECS's and are better than Random. 

For the \textit{IOF/ROL} dataset, it is evident that the results were inferior to \textit{Paint Control}, having small increments on performance as was expected from the analysis of Fig. \ref{rq1}.
Comparison methods appear to have slightly better performance during the first 200 CI cycles and afterwards RETECS seems to converge to a similar performance. This is an evidence of slow convergence, as it is clear in the second column of Fig. \ref{rq1}, and that more records are needed to possibly surpass the performance these methods.

For the \textit{Finance} data, there is clearly a learning pattern with an adaptation phase of around 90 commits which the RETECS method requires to have a similar performance to the traditional methods and a significant improvement with respect to Random. Then for the following commits, Random method progressively catches up with other methods, which can be a sign of a mutating environment, i.e. test cases at commit 300 are not failing for the same reasons that they were in commit 90. Overall, the algorithm achieves promising results, when applied to this novel dataset.

In conclusion, it is evident that RETECS can not perform significantly better than traditional methods. RETECS starts without any representation of the environment and it is not specifically programmed to pursue any given strategy. Yet it is possible to make prioritizations as good as traditional methods commonly used in the industry and by increasing the number of records available on each dataset, adding more features and conducting a more refined parameter tuning analysis, there is strong evidence that there can be a performance boost.

\subsection{Threats to Validity}
\subsubsection{\textbf{Internal}} 
RETECS is not a deterministic algorithm and because of its exploratory nature, randomness influences the outcomes. In order to mitigate the effect of random decisions, experiments are run for 30 iterations and the presented results correspond to an average of those results.  But it could be the case that a subsequent implementation of the method fails to produce the same results observed in this work. \par The second threat is related to parameter selection, that due to limited computer power, should have been more extensive. Thus, the chosen parameters are most likely not optimal for each scenario. Ideally, for each specific environment, parameters should be thoroughly adjusted.
Finally, due to the fact that this version of RETECS is an extension of the work developed by Spieker \textit{et al.} \cite{Spieker}, there's a threat related to implementation issues. Machine learning algorithms were implemented with the \textit{Scikit-Learn} library and the framework is available online for validation and reproducibility, in the aforementioned \textit{Github} repository. 

\subsubsection{\textbf{External}} 
The main gap related with external threats, pointed out by Spieker \textit{et al.} \cite{Spieker}, was the fact that the inclusion of only three datasets  was not representative of the wide diversity of CI environments. In the original paper three datasets are used, namely \textit{IOF/ROL}, \textit{PaintControl} and \textit{GSDTSR} from Google. The latter was not considered in this study due to its size and our inability to train it in feasible time. Although this study bridges this gap by including a novel dataset, increasing data availability and providing more validity to this framework, four datasets still fall short of a representative number of examples.

\subsubsection{\textbf{Construct}}

In this study, the concepts of failed test cases and faults are considered indistinguishable and interchangeable, yet this is not always true. For example, two test cases can fail due to the same fault, and, vice-versa, one failed test-case can reveal two or more faults. Nevertheless, because information about total faults is not easily accessible, the assumption that each test case indicates a different fault in the system is formulated. 
\par Regarding function approximators, there are many other machine learning algorithms that could be tested and fine-tuned to have a more accurate state space representation, steering the agent with more precision.
Regarding the information the agent uses in the decision process, (i.e. duration, last execution and execution history), it has proven to fail to surpass significantly the performance of simpler traditional methods, like \textit{Sorting}. To bridge this gap, more features should be added to enrich the information the agent has on each state, such as by using code-coverage, so that only test cases that will affect the files modified in a certain commit are considered.
Finally, RETECS was only compared to three baseline approaches, although there are more in the literature that should be considered, including other machine learning methods.

\subsection{Future Work}

The results obtained strongly indicate that RETECS can match performance with traditional prioritization methods and is flexible enough to adapt to different contexts. However due to its higher complexity in relation to traditional methods, unless its performance surpasses these other methods, it's not actually worth implementing. For its performance to increase, it requires more information to formulate better reasoning of expected failures, e.g. links between test cases and modified files. 
\par Regarding Machine Learning models as function approximators, DT's showed a slighty worse performance when compared to Network, however without collecting more records and a more refined parameter tuning analysis to try to find optimal values for all parameters, it is not possible to discard it. Additionally other models, rather than DT's, can and should be considered to represent memory, such as Nearest Neighbours.
\par Furthermore, in real world environments, test-cases are usually run on a grid that allows for parallelization. In our framework, we assumed that test cases were applied sequentially, one by one, based on their rank. If two test cases are very similar, most likely they will appear together in a test-schedule and detect exactly the same fault. With parallelization, it would be more fruitful to create groups of non-redundant tests to maximize the state-space covered by each group of test-cases on each run. 

\section{Conclusion}\label{conclusion}
 
 In this study, an extension of the RETECS framework was developed, in order to determine its ability to prioritize and select test-cases, when presented with a novel dataset, extracted from a different CI environment, validating its generalization. Additionally, DT's were applied for the first time in this context as a model for state space representation. \par
 Results indicate that RETECS can effectively create meaningful test schedules in different contexts. In the new \textit{Finance} dataset, with a combination of the Test Case Failure reward with the Artificial Neural Network Approximator, around 90 commits suffice to reach the performance of deterministic methods and surpass random prioritization of test-cases. Initially, the evaluation metric NAPFD starts at a value of only $0.2$ but, as the algorithm progresses, the trend shows values over $0.6$. \par
 The inclusion of DT's in the framework failed to produce better results relative to the Network, in the best possible case. However in some cases, with other reward functions, performance is comparable and might not be discarded right away, as it can be useful to apply, in future research, to other CI environments with distinct characteristics.


\bibliographystyle{unsrt}
\bibliography{bibliography.bib}

\end{document}